\documentclass[runningheads]{llncs}
\usepackage[T1]{fontenc}
\usepackage{subfigure}
\usepackage{graphicx}
\usepackage[hidelinks]{hyperref} 
\usepackage{color}

\usepackage{orcidlink}
\usepackage{cite}
\usepackage{enumitem}
\setlist{nosep}

\newcommand{\repthanks}[1]{\textsuperscript{\ref{#1}}}
\makeatletter
\patchcmd{\maketitle}
{\def\thanks}
{\let\repthanks\repthanksunskip\def\thanks}
{}{}
\patchcmd{\@maketitle}
{\def\thanks}
{\let\repthanks\@gobble\def\thanks}
{}{}
\newcommand\repthanksunskip[1]{\unskip{}}
\makeatother

\newcommand\add[1]{{#1}}

\begin{document}
	\title{Budget-Aware Keyboardless Interaction}
	\titlerunning{Budget-Aware Keyboardless Interaction}
	%
	\author{Quang-Thang Nguyen \thanks{Both authors contributed equally to this research.\protect\label{X}} \inst{1,2}\orcidlink{0009-0007-7374-6400}
		\and Gia-Phuc Song-Dong \repthanks{X} \inst{1,2}\orcidlink{0009-0009-3491-930X} 
		\and Minh-Triet Tran \inst{1,2}\orcidlink{0000-0003-3046-3041}
		\and Trung-Nghia Le \thanks{Corresponding author.} \inst{1,2}\orcidlink{0000-0002-7363-2610}
	}
	\authorrunning{Q.-T. Nguyen et al.}
	%
	\institute{University of Science, Ho Chi Minh city, Vietnam \and Vietnam National University, Ho Chi Minh city, Vietnam\\
		\email{\{22120333,22120282\}@student.hcmus.edu.vn},
		\email{\{tmtriet,ltnghia\}@fit.hcmus.edu.vn}
	}
	\maketitle              
	\begin{abstract}
		Interacting with computers typically relies on traditional input devices such as keyboards, mice, and monitors, which can be cumbersome for users seeking greater mobility. Virtual keyboards have been explored to address these limitations, but they often involve complex setups or expensive equipment. This paper proposes a novel virtual keyboard system that leverages only a standard camera and a paper with a printed keyboard layout. Unlike previous methods requiring complex calibration or special lighting conditions, our approach can work on standard environment using modern computer vision technologies. Combining modern segmentation and detection models with traditional image processing algorithms, we efficiently identify the keyboard region. Touch detection is performed using an algorithm analyzing the color of the user’s fingernail. Experiments demonstrated a promising results our proposed solution of keyboard and keystroke detection for practical applications. Participants attended our user study also found the proposed system interesting.
	\end{abstract}
	\section{Introduction}
	
	\begin{figure}[t!]
		\centering
		\subfigure[Camera and keyboard setting up]{
			\includegraphics[width=0.45\columnwidth]{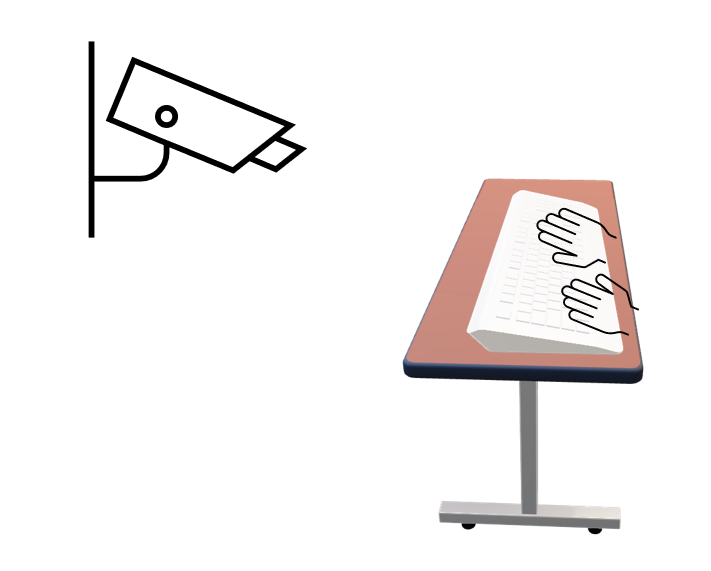}
			\label{figure:set up camera and keyboard}
		}
		\subfigure[Keyboard-printed paper]{
			\includegraphics[width=0.45\columnwidth]{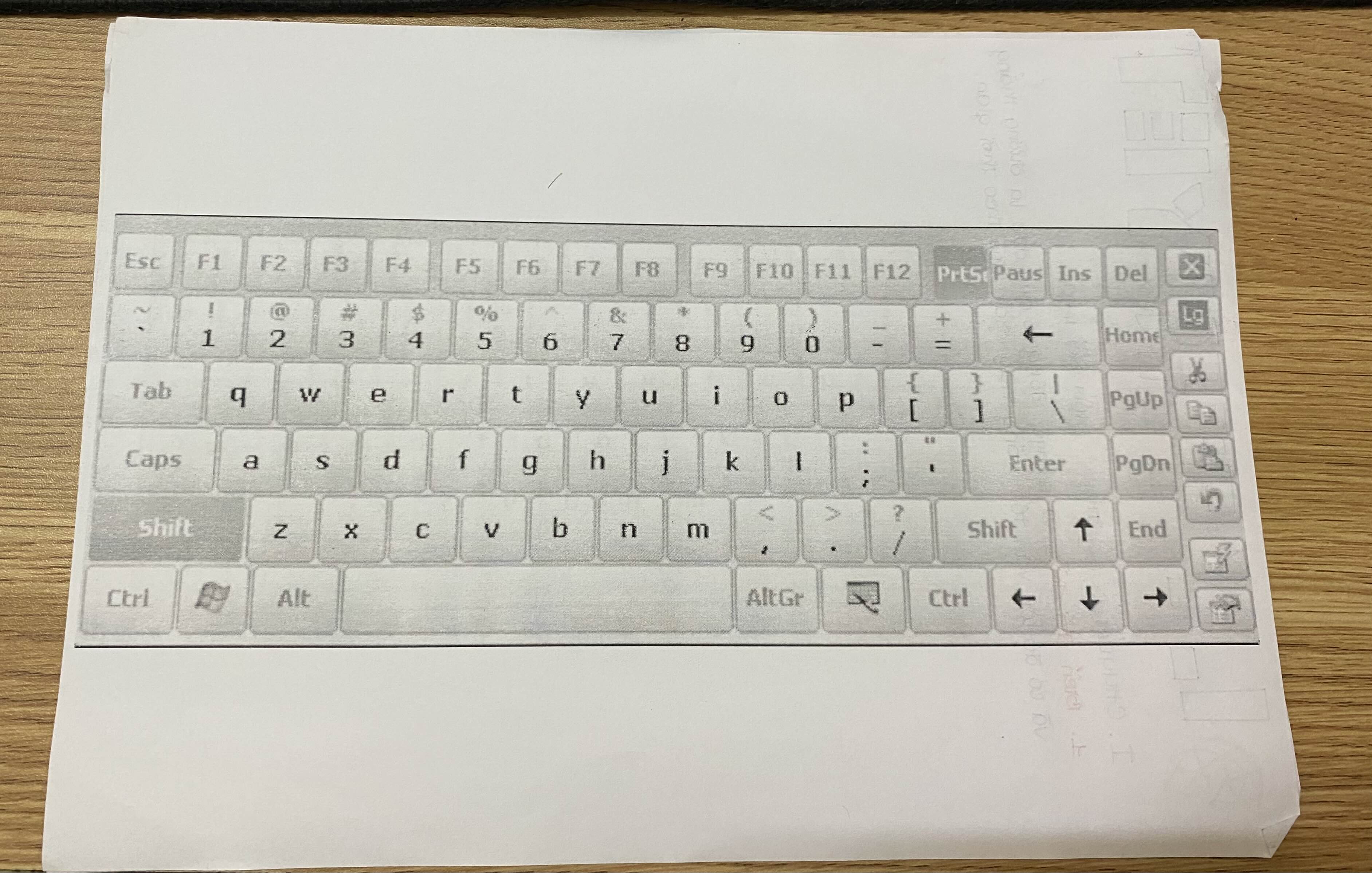}
			\label{figure:paper keyboard}
		}
		\caption{Setup of the proposed system, which leverages only a standard camera and a keyboard image without any markers.}
		\label{fig:hardware requirements}
	\end{figure}
	
	Traditionally, interacting with computers has depended on keyboards, mice, and monitors, which can be cumbersome and inconvenient for users requiring greater mobility. To address this challenge, various innovative methods have been proposed. For instance, virtual keyboards can now be utilized with flat surfaces \cite{Virtual_keyboard_projector_3D_camera_Du2005,Virtual_keyboard_phone_standard_camera_Posner2012}, or in AR/XR environments \cite{Text_input_QwertyRing_Gu2020,Virtual_keyboard_AR_Yıldıran2022,Virtual_keyboard_XR_Shatilov2023}. The TapType \cite{Text_input_TapType_Streli2022} system enables users to type using two wristbands, eliminating the need for a physical keyboard. Yıldıran \cite{Virtual_keyboard_AR_Yıldıran2022} uses virtual hands and a virtual keyboard in an XR environment, and allows users to adjust the keyboard's size.
	
	There are two primary types of keyboardless input approaches: Dynamic Bayesian Networks \cite{Virtual_keyboard_XR_Fu2024,Text_input_QwertyRing_Gu2020,Text_input_TapType_Streli2022} and projected virtual keyboards \cite{Virtual_keyboard_phone_standard_camera_Posner2012,Virtual_keyboard_projector_3D_camera_Du2005,Virtual_keyboard_AR_Yıldıran2022,Virtual_keyboard_XR_Shatilov2023}. Dynamic Bayesian Networks may struggle with non-standard typing patterns since they guess keystroke by analyzing finger's position when typing. The same position of finger returns different key in different keyboard layout. Paper interaction system proposed by Adajania et al. \cite{Virtual_keyboard_paper_standard_camera_Adajania2010} requires markers printed on keyboard images that constrain the accessibility of the system. Meanwhile, projected virtual keyboards often require bulky or expensive equipment such as projectors or AR headsets. An AR headset is priced around 400 USD, while a projector costs 200 USD, and both are cumbersome to transport and use. Therefore, these systems are difficult to deploy for practical use.
	
	In this paper, we introduce a novel approach for virtual keyboard using only a standard camera and a sheet of paper with a printed keyboard layout. In our system, a camera required to record the typing user is readily available on any laptop or smartphone, therefore it is much cheaper than existing systems relying on multiple cameras \cite{Touch_detection_3d_Katz2007} or special camera \cite{Virtual_keyboard_projector_3D_camera_Du2005,Touch_detection_highspeed_camera_Yamamoto2006,Touch_detection_stereo_camera_Agarwal2007}. Our proposed system supports a variety of keyboard layouts, including QWERTY and AZERTY, as well as keyboards from MacOS, like the Magic Keyboard, and those from Windows. Specially, unlike the virtual keyboard proposed by Adajania et al. \cite{Virtual_keyboard_paper_standard_camera_Adajania2010}, we do not need to mark anything on the printed image.  
	
	Figure~\ref{figure:set up camera and keyboard} shows the setup of our proposed virtual keyboard system. A standard camera is position in a frontal, slightly angled view, pointing down towards the keyboard area. The system can be easily run on a personal computer with a webcam. By detecting the keyboard region beforehand, users can type on the printed keyboard without any hands in the frame, allowing real-time practical deployment. 
	
	The proposed system consists of two main modules: the keyboard processing module identifies the keyboard region and its keys and the touch processing module detects whether a touch is made. YOLOv8n-seg and YOLOv8n \cite{yolov8_ultralytics} are utilized for keyboard and keystroke detection. The printed keyboard is transformed to orthogonal view using homography transformation \cite{Homography_Babbar2022} to improve the detection performance. On the other hand, YOLOv8n-seg \cite{yolov8_ultralytics} is used to segment fingernails for color analysis, followed by a color analysis algorithm \cite{Touch_detection_color_fingertips_Marshall2008} to identify whether fingers are pressing. \add{The users have to apply a moderate amount of pressure while pressing key to ensure the method works.} To expedite the process, we  analyze fingertip movement using the Google Mediapipe library \cite{Model_Mediapipe} and proceed with segmentation only for suspected candidates.
	
	We conducted extensive experiments and user study to analyzing the proposed system. The experimental results show that we achieved high AP of 92\% for keyboard detection and 70\% for detecting keys on that keyboard. The touch detection accuracy is only around 36\% due to affects of light intensity in the experiments. The results of the user study show that the majority of participants found our proposed system interesting, as they can input text with just a piece of paper and a camera.
	
	Our contributions are as follows:
	\begin{itemize}
		\item We present a novel method for constructing a virtual keyboard using a sheet of paper that has a printed keyboard layout, all without the need for additional markings.
		\item We train YOLO models to segment keyboard region and detect keystrokes.
		\item We employ the fingertip color analysis to detect touch utilizing only a single standard camera.
	\end{itemize}
	
	%
	%
	%
	\section{Related work}
	\label{section: Related work}
	\subsection{Virtual Keyboard}
	Adajania et al. \cite{Virtual_keyboard_paper_standard_camera_Adajania2010} used a paper sheet with a printed keyboard where the endpoints were highlighted in blue, aiding in their identification during thresholding and keyboard recognition. Posner et al. \cite{Virtual_keyboard_phone_standard_camera_Posner2012} developed a virtual floating keyboard for mobile phones, designed to function on any surface by utilizing the phone's single 2D camera to capture and interact with the keyboard. Du et al. \cite{Virtual_keyboard_projector_3D_camera_Du2005} employed a projector to display the virtual keyboard on a surface. Yildiran et al. \cite{Virtual_keyboard_AR_Yıldıran2022} employed AR/VR Head-Mounted Displays (HMDs) to render a virtual keyboard and two virtual hands within the display, allowing users to interact with the keyboard entirely in a virtual environment. Shatilov et al. \cite{Virtual_keyboard_XR_Shatilov2023} introduced MyoKey, a text entry system for XR headsets that relies on inertial measurement units (IMUs) and myoelectric signals instead of virtual objects, thereby reducing the number of gestures needed and improving text input efficiency.
	
	With the advancement of deep learning, methods that enable typing without relying on physical or virtual keyboards began to emerge in the 2020s. Fu and Xi \cite{Virtual_keyboard_XR_Fu2024} introduced a deep-learning approach to infer keystrokes from media channels captured by AR headsets. Concurrently, Gu et al. \cite{Text_input_QwertyRing_Gu2020} developed QwertyRing, a text entry technique that utilizes signals from a finger-worn 6-axis IMU along with a Bayesian decoder, making it compatible with external displays such as AR/VR headsets or smart TVs. Additionally, the Taptype system by Streli et al. \cite{Text_input_TapType_Streli2022} enabled typing using inertial sensors embedded in a wristband, allowing users to type without needing to remove their mobile phones from their pockets. 
	
	Different from existing systems, our proposed system employs a printed keyboard layout on a single sheet of paper, which can be in black and white or color, without requiring any extra markings. For keystroke inference, a cheap camera on any commercial laptop or mobile phone is sufficient to capture both the hands typing and the keyboard, resulting in a highly cost-efficient virtual keyboard system.
	
	\subsection{Touch Detection}
	Touch detection could be achieved based on depth sensors or stereo cameras. Yamamoto et al. \cite{Touch_detection_highspeed_camera_Yamamoto2006} utilized a high-speed vision camera to concentrate on the high-frequency component that arose when a fingertip contacted an object. Du et al. \cite{Virtual_keyboard_projector_3D_camera_Du2005} proposed a 3D optical ranging-based virtual keyboard system that uses a pattern projector and a 3D range camera. Their system reconstructs typing events by analyzing both gray-scale and depth information from the scene, specifically examining the depth curve of the finger to detect touch. With an overhead camera and a side-mounted camera, Katz et al. \cite{Touch_detection_3d_Katz2007} was also able to calculate the three-dimensional coordinates of the fingertips and the surface. Agarwal et al. \cite{Touch_detection_stereo_camera_Agarwal2007} introduced a computer vision algorithm that uses an overhead stereo camera to detect touch with high precision by aggregating stereo cues from several fingertip points. However, these methods necessitate special or multiple cameras, leading to higher costs or greater complexity during setup.
	
	Meanwhile, shadow analysis algorithms were represented. Song et al. \cite{Touch_detection_projector_Song2007} used one camera and one projector overlooking the tabletop. When two fingertips of the real finger and its shadow created by the projector merged into one, a touch was detected. Posner et al. \cite{Virtual_keyboard_phone_standard_camera_Posner2012} used the similar idea to create their virtual keyboard by detecting the fingertip and its shadow's tip using environment light and a standard 2D camera.
	Also based on shadow analysis, Adajania et al. \cite{Virtual_keyboard_paper_standard_camera_Adajania2010} proposed the system that detects touch by analyzing the ratio of white pixels, which represent areas not covered by shadows, to black pixels, which represent shadowed areas. Although these approaches required no additional hardware beyond the web camera, their sensitivity to lighting direction can sometimes lead to insufficient shadow changes, making touch detection less reliable.
	
	On the other hand, Marshall et al. \cite{Touch_detection_color_fingertips_Marshall2008} developed a method to detect finger pressure by analyzing color changes in the fingertip, which occurred due to blood displacement when pressing against a hard surface. Their approach employed a standard camera and computer to visually capture these color variations, allowing for pressure and multi-touch sensing on diverse surfaces without altering the physical object. Inspired by the work of Marshall et al. \cite{Touch_detection_color_fingertips_Marshall2008}, we improve their algorithm by excluding the normalization step to speed up the nail color analysis.
	
	\section{Proposed System}
	\label{section:Proposed System}
	
	\subsection{Overview}
	
	
	Our system is represented in Figure~\ref{figure:set up camera and keyboard}, where the camera is positioned in front and angled upward, so that it clearly captures the keyboard and the hands typing on it. \add{The camera angle should not be too vertical, as it will make it difficult to observe the nails. At the same time, it should not be too low, as this may impair the ability to recognize the keyboard (Section~\ref{section: Experiment}). It should be between $45^\circ$ and $60^\circ$.} We use a sheet of paper with a printed keyboard design instead of a physical keyboard or a virtual keyboard projected by a projector. The image can be printed in color or black and white. The size should be large enough to comfortably type by hand (i.e., A4 paper). Only one standard camera is needed, and this can be found on any laptop or smartphone. \add{When pressing a key, users need to apply sufficient pressure for the color of the fingernail to change noticeably, allowing the system to function properly.}
	
\begin{figure}[t!]
		\centering
		\includegraphics[width=\linewidth]{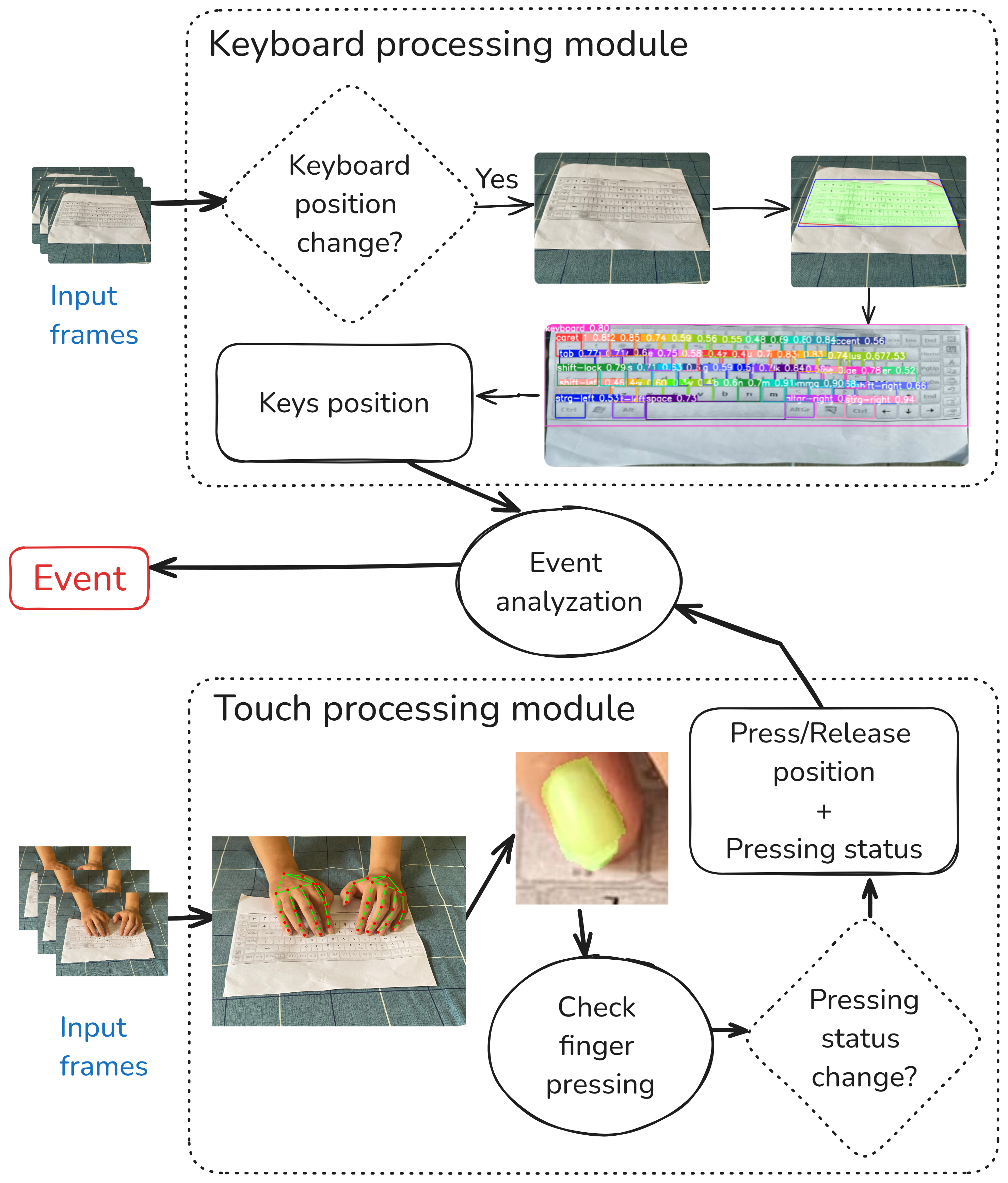}
		\caption{Flowchart of the proposed keyboardless interaction system.}
		\label{figure:system architecture}
	\end{figure}

	Our system consists of two primary modules: Keyboard processing module and Touch processing module, as illustrated in Figure~\ref{figure:system architecture}. The keyboard processing module identifies the bounding boxes of the keys. Meanwhile, the touch processing module detects the location where the finger touches or releases the key. This position is then used to determine which key has been pressed or released.
	
	\subsection{Keyboard processing module}
	\label{subsection: keyboard processing module}
	
	\begin{figure}[t!]
		\centering
		\includegraphics[width=\columnwidth]{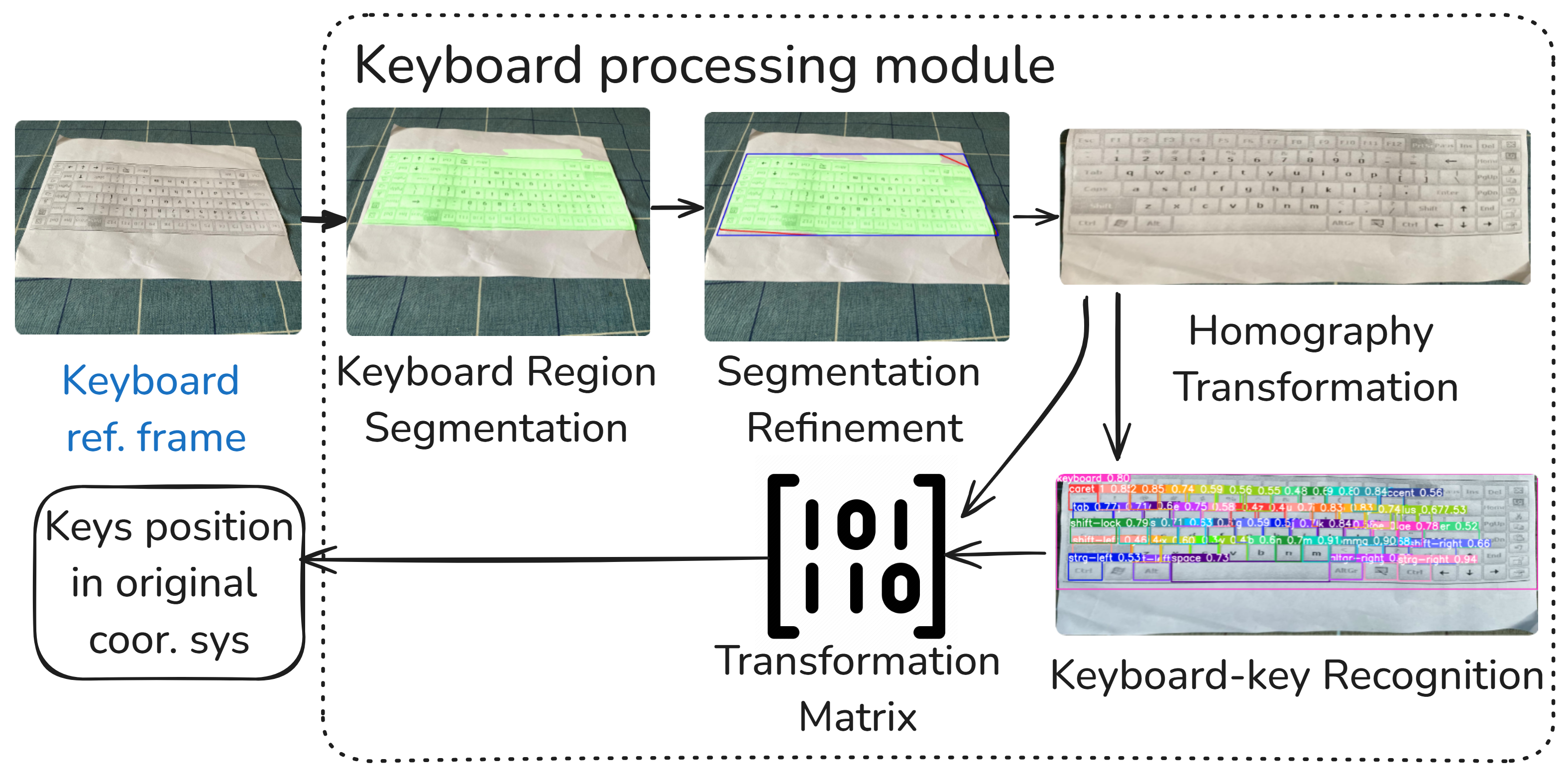}
		\caption{Pipeline of keyboard processing module.}
		\label{fig:keyboard processing module diagram}
	\end{figure}
	
	Figure~\ref{fig:keyboard processing module diagram} illustrates overview pipeline of keyboard processing module. First, we draw a quadrilateral that closely surrounds the edges of the keyboard. Subsequently, the keyboard area is transformed to a top-down orthogonal view using homography transformation. Finally, the positions of the keys on the keyboard are identified utilizing a finely-tuned YOLO model \citeonline{yolov8_ultralytics}. The details of the steps are presented as follows.
	
	\subsubsection{Keyboard Region Segmentation. }
	
	We utilize a semantic segmentation model to locate the positions of keyboards within image frames. Prior to processing, each image frame is converted to gray-scale, as color can introduce noise, especially if keyboards are inconsistently colored. The processed frames are then fed into the model to segment the keyboards. The results are stored as coordinate masks for each keyboard present in the frame.
	
	\textbf{\textit{Implementation: }} We employ the YOLOv8s-seg.pt pre-trained model and train it for 30 epochs. For fine-tuning, we use one label (keyboard). The learning rate is set at 0.01, and the batch size is 16.
	
	\subsubsection{Segmentation Refinement. }
	
	With a top-down and angled view, the keyboard image is not rectangular but trapezoidal. This makes it difficult for the model to recognize the keys' position on the keyboard. Therefore, we use the convex hull of these points to create a convex polygon, which maps the image back to a perpendicular projection angle.	Furthermore, we construct a quadrilateral encompassing the keyboard area by utilizing the Minimum Area Enclosing Polygon algorithm \cite{Minimun_convex_k-gons_Aggarwal1985}.
	
	\subsubsection{Homography Transformation. }
	
	Once the four vertices of the keyboard are identified, we employ Homography \cite{Homography_Babbar2022} to transform the image from an oblique perspective into a rectangular keyboard image with a top-down orthogonal view. With this perspective, recognizing individual keys on the keyboard becomes simpler. Particularly, characters on the keys undergo minimal distortion, facilitating the key identification process as bounding boxes do not overlap as much as in other viewing angles. Additionally, each point pressed by the user's hand corresponds to a unique bounding box. This enables to swiftly determine which key is pressed, optimizing pressed key search.
	
	\subsubsection{Keyboard-key Recognition. }
	
\begin{figure}[t!]
		\centering
		\subfigure{
			\includegraphics[width=0.45\columnwidth]{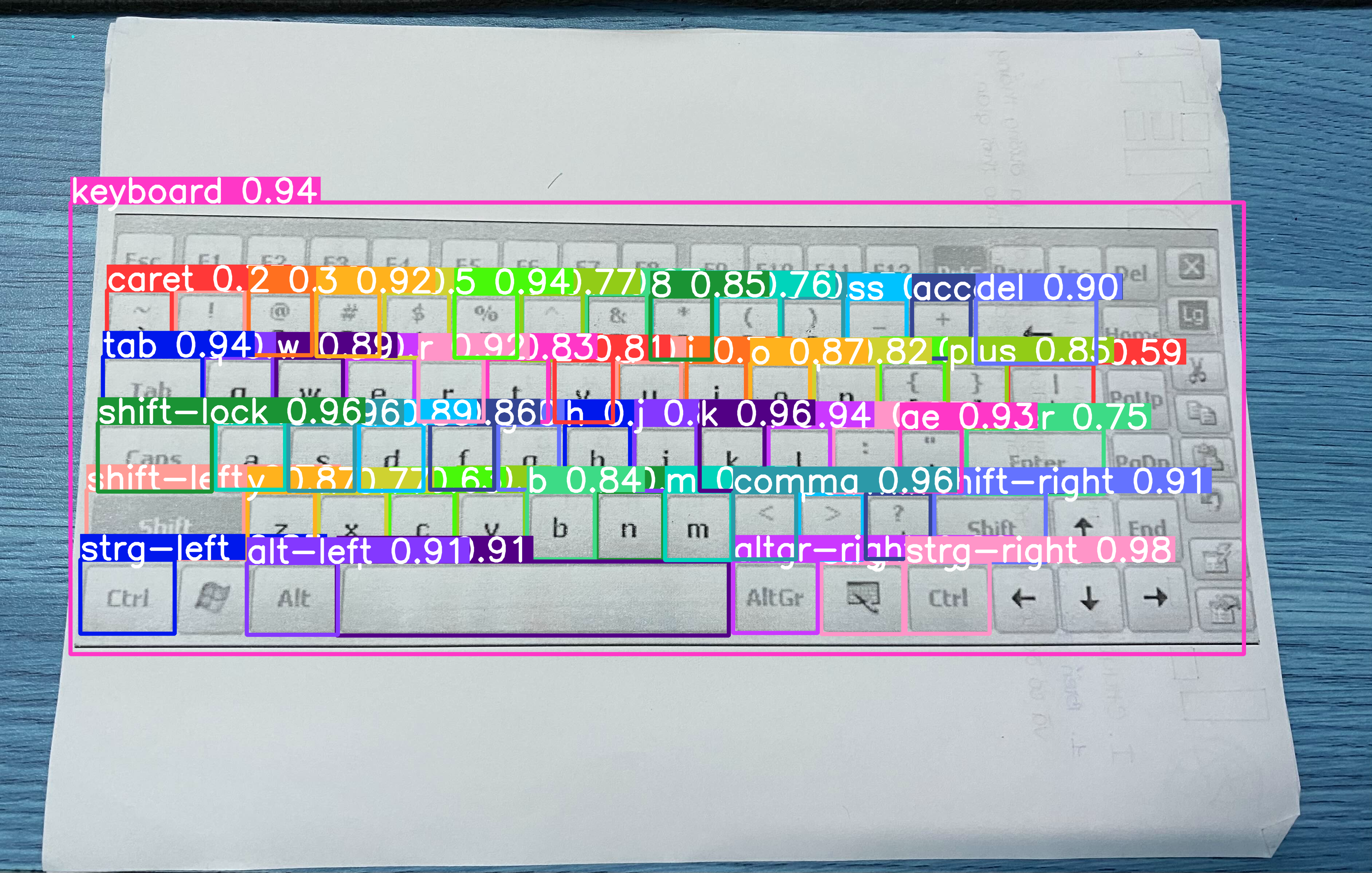}
		}
		\subfigure{
			\includegraphics[width=0.45\columnwidth]{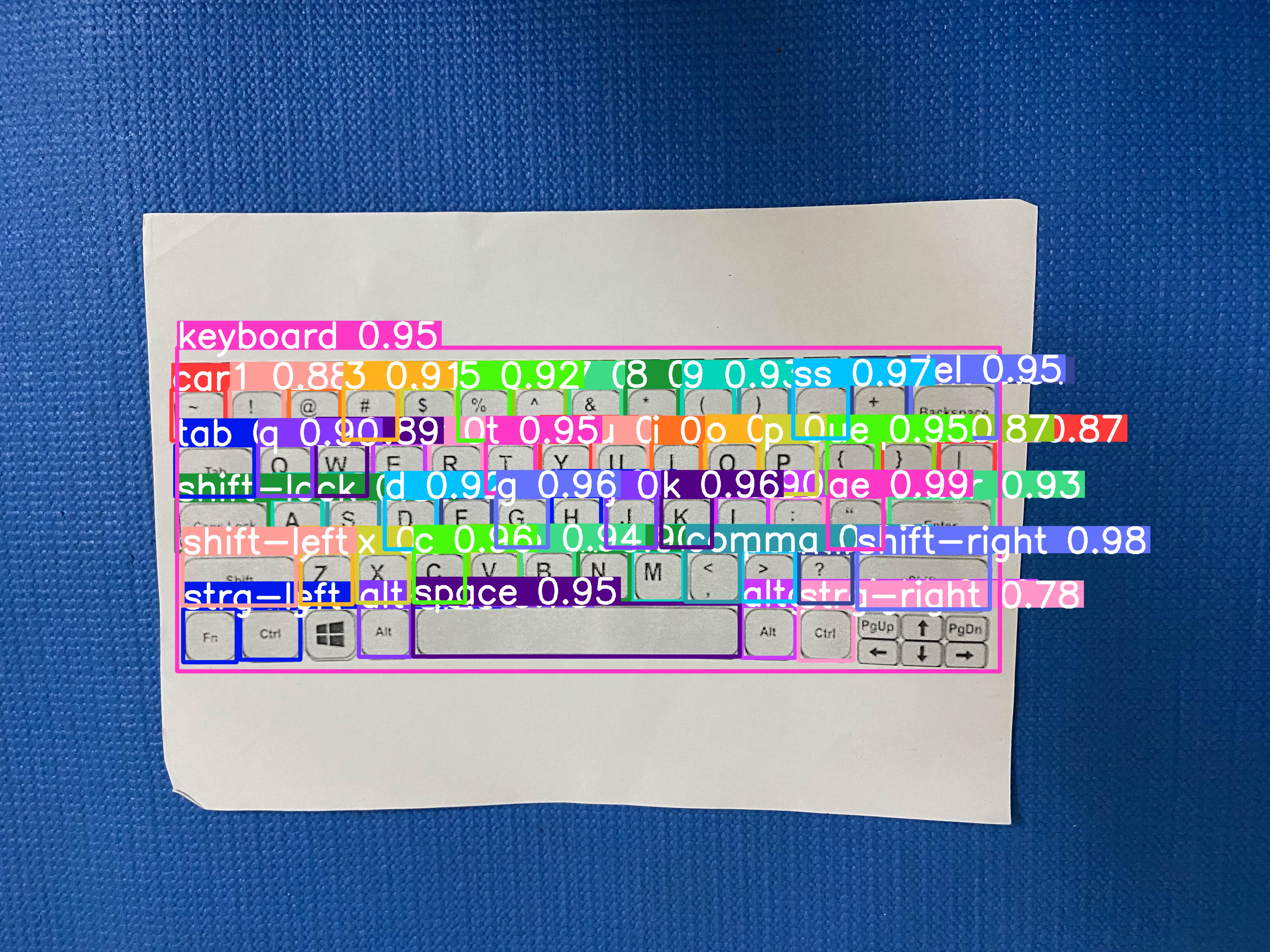}
		}
		\caption{Keyboard with keys are detected and marked.}
		\label{figure:Keyboard key recognition}
	\end{figure}

	We utilize an object detection model to pinpoint the positions of individual keys on the keyboard, resulting in a clearly delineated keyboard with each key's location distinctly marked, avoiding overlap (see Figure~\ref{figure:Keyboard key recognition}).
	
	\textbf{\textit{Implementation: }} We utilize the pre-trained YOLOv8n.pt model and train it for 18 epochs. Our fine-tuning process involves 60 labels, corresponding to the 60 keys on a standard keyboard. We employ a batch size of 16, with a learning rate set at 0.01 to optimize the training process.
	
	\subsection{Touch processing module}
	
	\begin{figure}[t!]
		\centering
		\includegraphics[width=\linewidth]{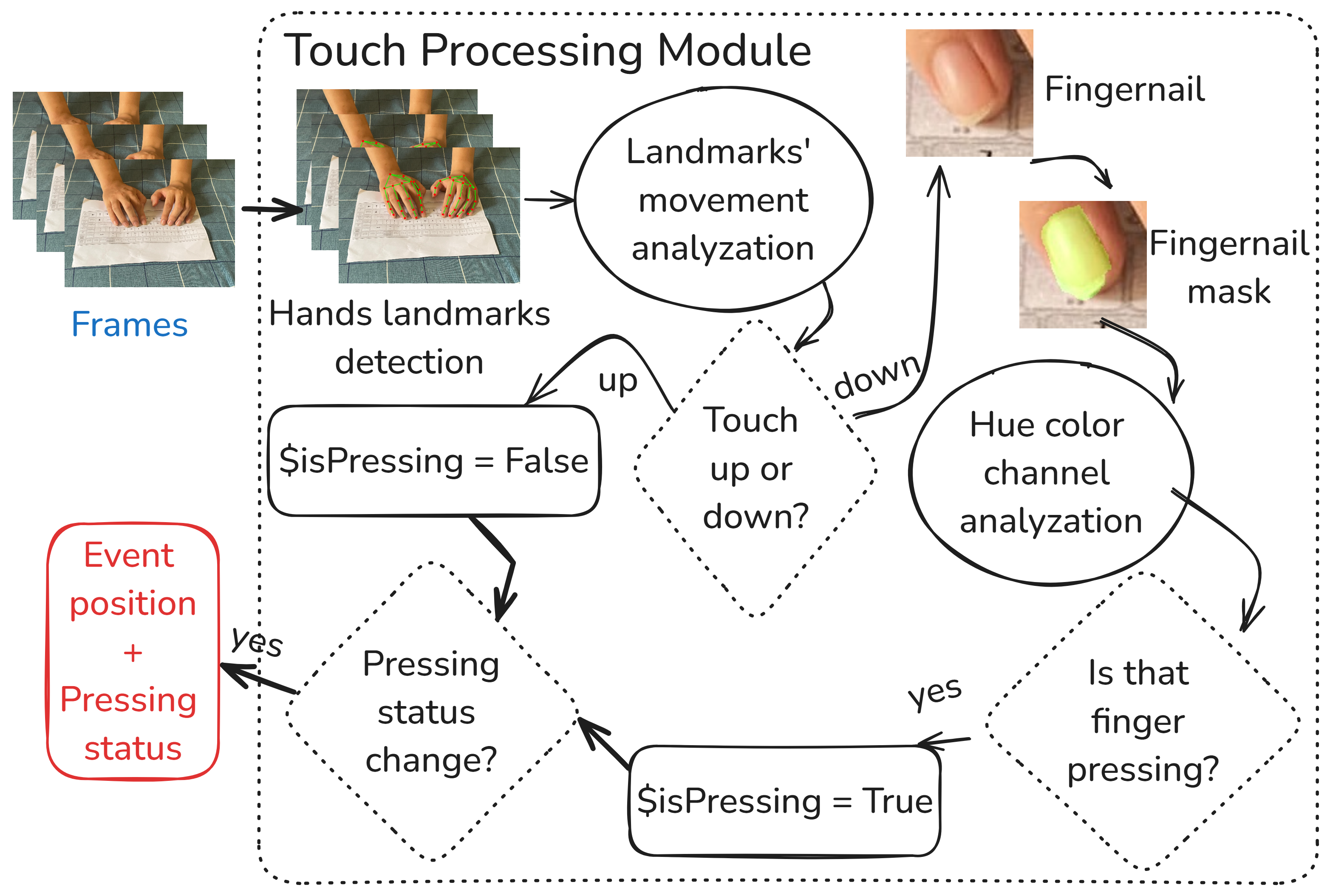}
		\caption{Pipeline of touch processing module.}
		\label{figure:touch processing module}
	\end{figure}
	
	Figure~\ref{figure:touch processing module} illustrates the pipeline for touch detection. First, the video frame is passed through the hand landmarks detection model~\cite{Model_Mediapipe} to determine the position of the fingertips . Then we compare position of the fingertips within a sufficiently small neighborhood of current frame and return one of these event: press a key or keep holding a key (i.e., touch down), release a key (i.e., touch up), and press nothing (i.e., finger is in the air).
	
	With the touch up event, we set pressing status to False, and jump to the last step of this module. With the key down event, we analyze color of the fingertips to make sure there is a touch down event with higher confident level.
	
	To determine whether the finger is pressing a key, we segment the fingernail and analyze its color. Particularly, we extract a small image region whose center at the position detected by the hand landmarks detection model and has a predefined size. This frame is then passed through the segmentation model to detect the actual nail area. Similar to the work of Marshall el al. \cite{Touch_detection_color_fingertips_Marshall2008}, we calculate Hue values as follows: 	
	\begin{equation}
		MeanHue = \arctan \left( \left[ \sum_1^n \cos (Hue) \right], \left[ \sum_1^n \sin (Hue) \right] \right),
	\end{equation}
	\begin{equation}
		VarHue = \frac{1}{n} \sum_1^n \min \left( (Hue - MeanHue)^2, (360 - (Hue - MeanHue))^2 \right),
	\end{equation}
	
	Base on the $VarHue$ value, we check whether this is indeed a press event or not. If there is a Touch down event (i.e., a finger is pressing a key), we set pressing status to True and move to last step.
	
	Finally, the pressing status is check whether changed or not. If it did, the user had just pressed or released a key. We combine the coordinate of this event with data from keyboard processing module to determine which key is pressed or release.
	
	To avoid excessive segmentation of the fingernail, which is a slow process, we do a calculation to find out if there is a touch down or touch up with low confident level or not. Let $|\Delta x|$ and $|\Delta y|$ represent the changes in the position of the fingertip in the horizontal and vertical directions, respectively. We checks how the new position differs from the positions of previous frames via four cases:
	\begin{itemize}
		\item If $|\Delta x|$ is too big, there might be something wrong in the hand landmarks detection model (e.g., the finger is covered by the others). We assumes the finger just release the key if it is pressing a key.
		
		\item If $\Delta y > 0$ and the finger is not pressing any key, it might be about to press a key. We note that and wait for next frame to know the finger keeps moving downward or stop when meeting a keyboard surface.
		
		\item If $\Delta y < 0$ and the finger is pressing a key, it might be about to release a key.
		
		\item If $|\Delta y| < \varepsilon$, where $\varepsilon$ is the movement threshold, we assumes that different is caused by the unstable when detecting hand landmarks, and the finger is considered not moving. We now use the color processing layer to determine whether there is a press event or not.
		
		\item In other cases, nothing changes. We go to the next video frame.
	\end{itemize}
	
	\section{Experiment}
	\label{section: Experiment}
	
	\subsection{Experimental Settings}
	
	We set up our system in natural light environment for experiments because it is difficult to observe the color of the nail polish under artificial light. A back camera in iPhone 13 Pro Max was used to record typing videos.

    \add{Before typing, users should leave the keyboard stationary on a flat surface for a moment to allow the system to detect the positions of the keys. Next, they need to position their fingers in front of the camera without pressing any keys, enabling the system to capture the initial color of their fingers.}
	
	For keyboard region detection, we utilized 2,007 images for training and 388 images for validation. The dataset is aggregated from various sources, such as Microsoft COCO \cite{Dataset_COCO_microsoft2024} (1,790 images for training and 388 images for validation), E-waste \cite{Dataset_e-waste-detection-model_TRCProject2023} (130 images for training), and MieKeyboard \cite{Dataset_keyboard-v2itg_MieUniversity2023} (130 images for training). For keyboard-key recognition, we used Microsoft COCO \cite{Dataset_COCO_microsoft2024} dataset for training and validating. For fingernail segmentation, we used dataset from KaggleNail \cite{Dataset_nail} (2,364 images for training, 913 for validating, 898 for testing).
	
	\subsection{Experimental Results}
	
	\subsubsection{Keyboard-key Detection. }
	
	\begin{figure}[t!]
		\centering
		\subfigure[Windows AZERTY keyboard with $30^\circ$ view]{
			\includegraphics[width=0.40\columnwidth]{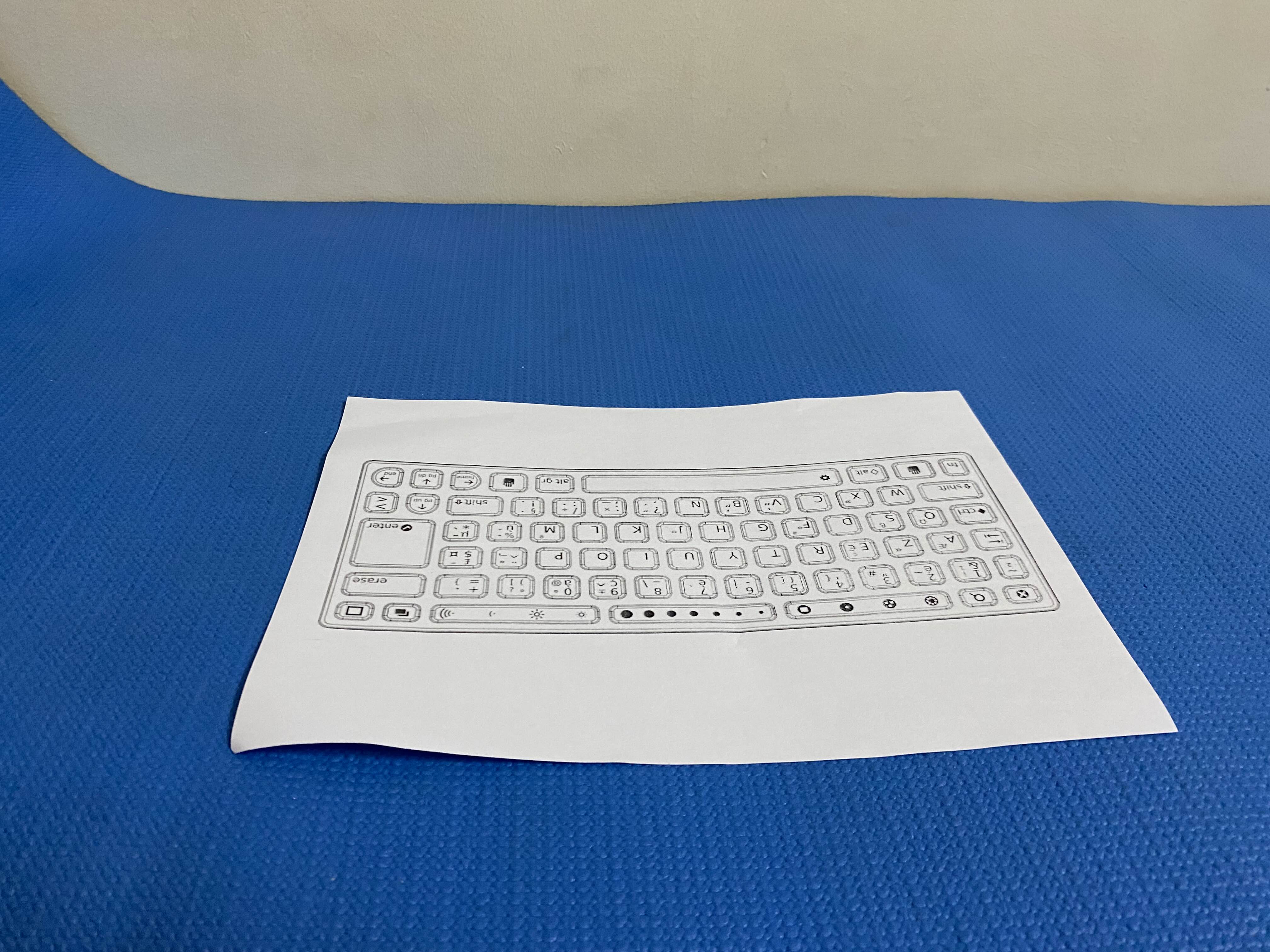}
			\includegraphics[width=0.55\columnwidth]{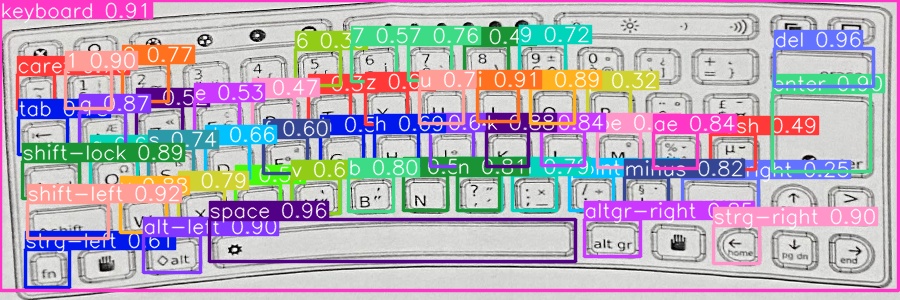}
		}
		\subfigure[Windows QWERTY $60^\circ$ view and Apple QWERTY $90^\circ$ view]{
			\includegraphics[width=0.41\columnwidth]{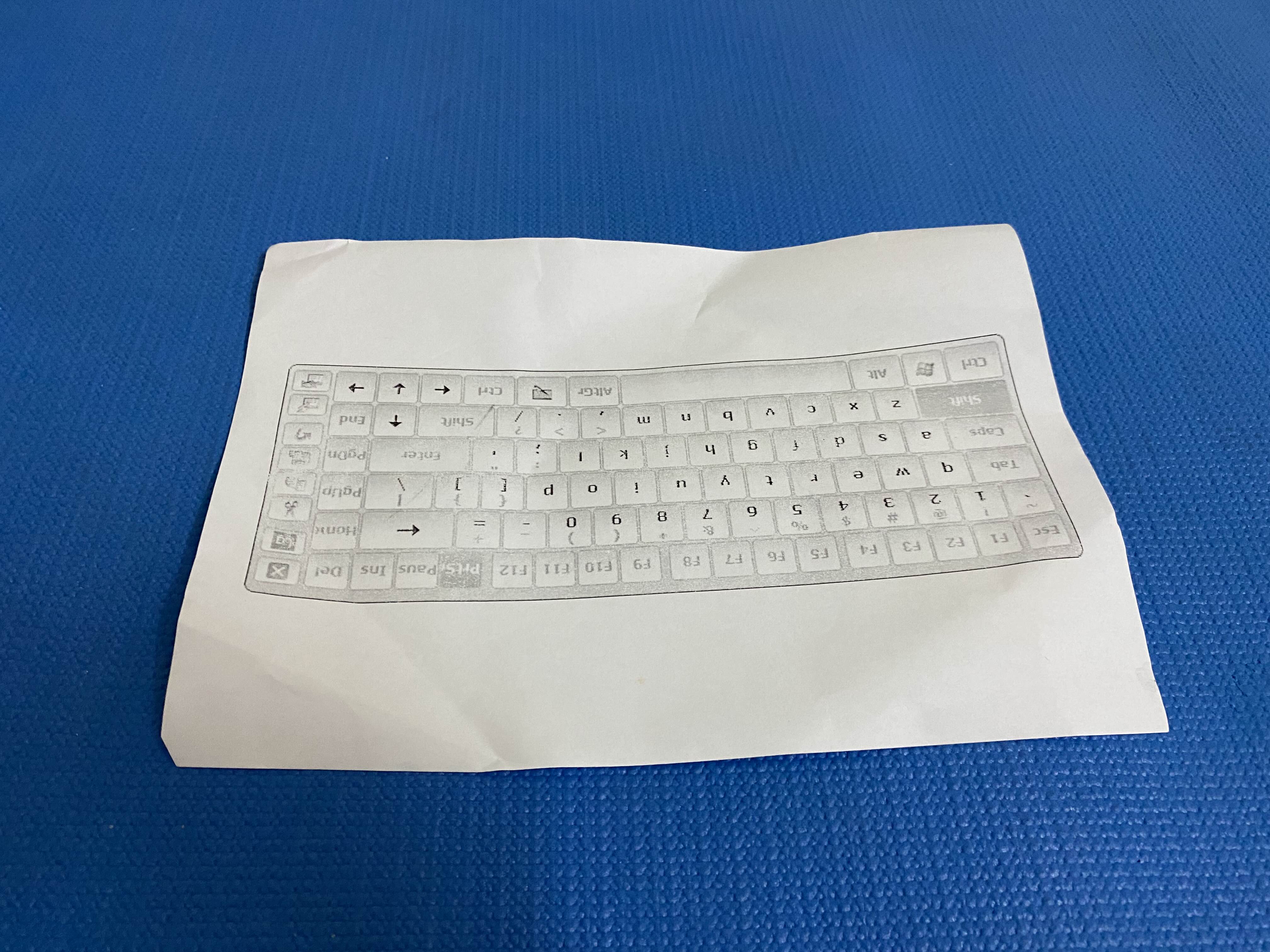}
			\includegraphics[width=0.41\columnwidth]{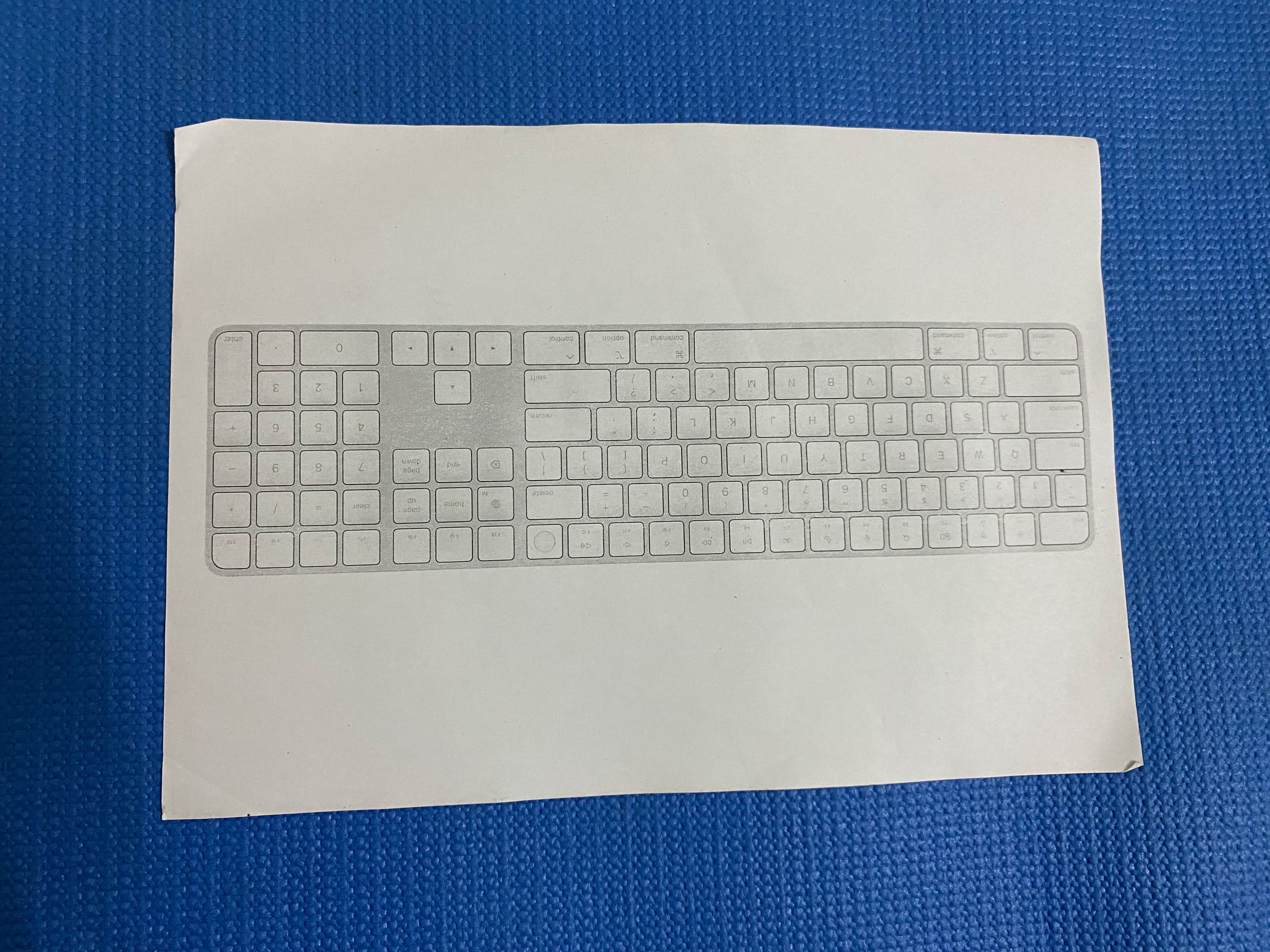}
		}		
		\caption{Illustration of keyboard layouts from different perspectives.}
		\label{figure:Keyboard processing module experiment}
	\end{figure}
	
	
	We took 12 photos of keyboard 3 with frontal view, in three different angles, as illustrated in Figure~\ref{figure:Keyboard processing module experiment}. Regardless captured images were not clear, keyboard region was detected with acceptable AP around $90\%$, as showed in Table~\ref{tab:keyboard processing module result}. Average accuracy of keys on keyboard is about 70\%. This detection performance is not so high because the dataset contains many keyboard layouts (e.g., QWERTY, AZERTY, Windows keyboard layout, Mac keyboard layout).
	
	\begin{table}[t!]
		\caption{Keyboard detection results.}
		\label{tab:keyboard processing module result}
  \centering
  \resizebox{\textwidth}{!}{%
				\begin{tabular}{|l|c|c|c|}
					\hline
					\multicolumn{1}{|c|}{\textbf{Type of keyboard}}                                 & \textbf{View angle} & \textbf{AP of   "keyboard" class} & \textbf{Average accuracy of keys} \\ \hline
					Mac QWERTY                                                                      & Frontal             & 0.95                              & 0.8                               \\
					Mac QWERTY                                                                      & 60                  & 0.95                              & 0.67                              \\
					Mac QWERTY                                                                      & 45                  & 0.94                              & 0.7                               \\
					Mac QWERTY                                                                      & 30                  & 0.96                              & 0.77                              \\ \hline
					Windows AZERTY                                                                  & Frontal             & 0.89                              & 0.64                              \\
					Windows AZERTY                                                                  & 60                  & 0.89                              & 0.64                              \\
					Windows AZERTY                                                                  & 45                  & 0.88                              & 0.57                              \\
					Windows AZERTY                                                                  & 30                  & 0.91                              & 0.69                              \\ \hline
					Windows QWERTY                                                                  & Frontal             & 0.91                              & 0.81                              \\
					Windows QWERTY                                                                  & 60                  & 0.93                              & 0.71                              \\
					Windows QWERTY                                                                  & 45                  & 0.92                              & 0.71                              \\
					Windows QWERTY                                                                  & 30                  & x                                 & x                                 \\ \hline
				\end{tabular}
    }
	\end{table}
	
	\subsubsection{Nail Segmentation. }
	
	\begin{figure}[t!]
		\centering
		\subfigure{
			\includegraphics[width=0.40\columnwidth]{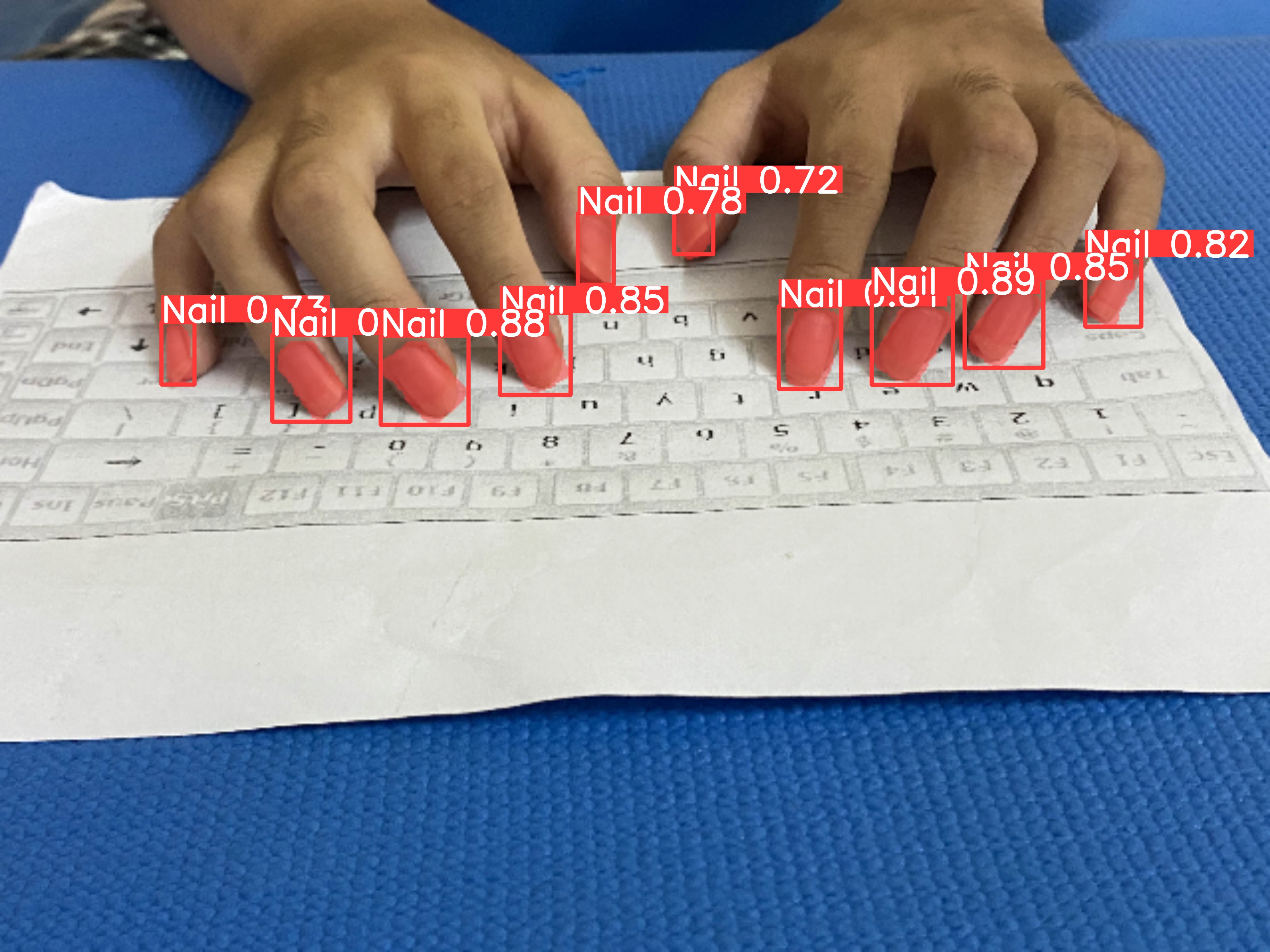}
		}
		\subfigure{
			\includegraphics[width=0.40\columnwidth]{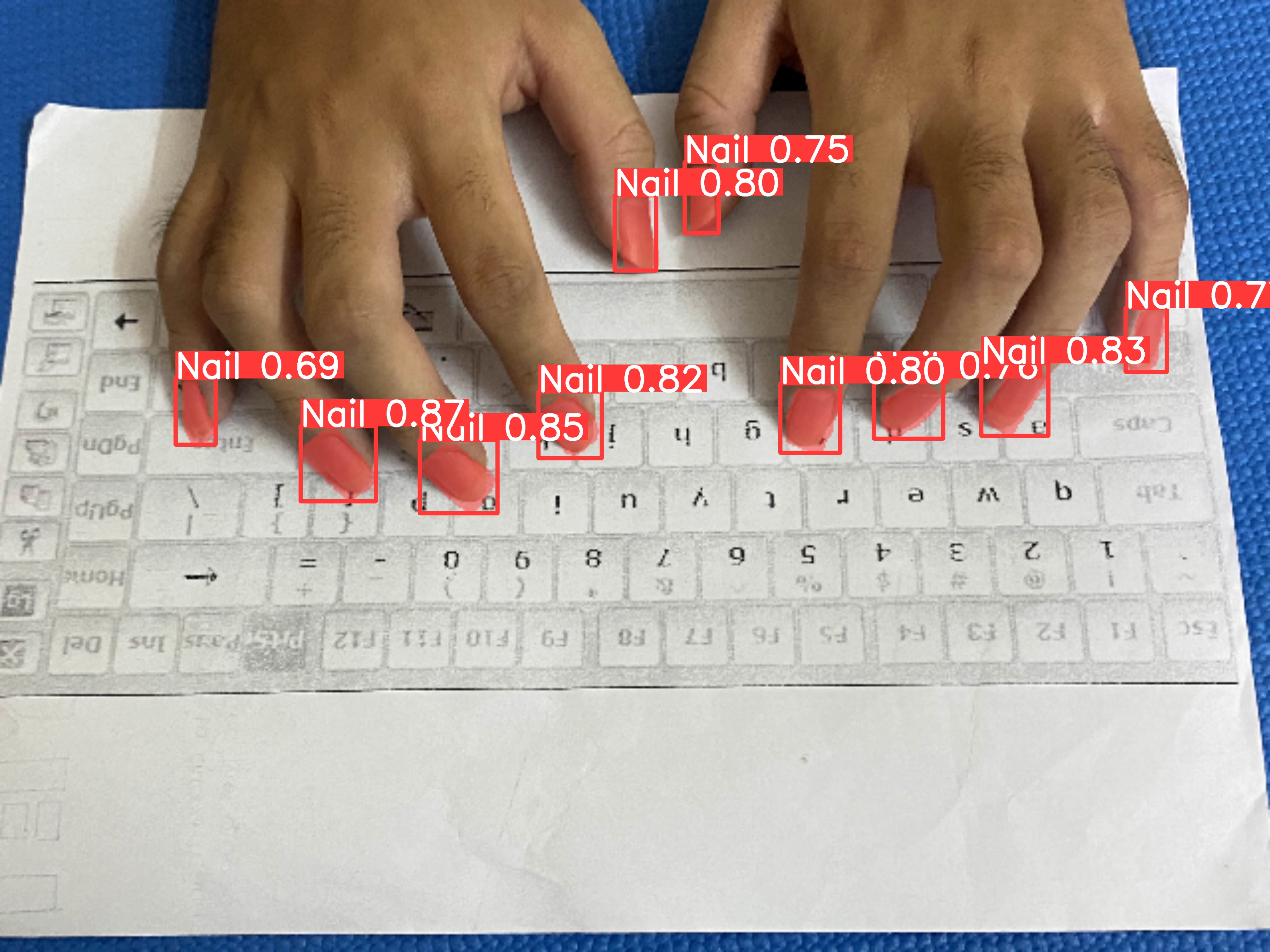}
		}		
		\caption{Hands with fingers segmented.}
		\label{figure:Hands with fingers segmented}
	\end{figure}
	
	We captured two images of typing hands from different angles, as shown in Figure~\ref{figure:Hands with fingers segmented}. The result is represented in Table~\ref{tab:nail segmentation result}. The thumbs and little fingers often have a lower recognition rate because their angles are usually not aligned with the camera angle.
	
	\begin{table}[t!]
		\caption{Nail segmentation result.}
		\label{tab:nail segmentation result}
		\begin{minipage}{0.8\columnwidth}
			\begin{center}
				\begin{tabular}{|c|c|c|c|c|}
					\hline
					\textbf{View angle} & \textbf{Mean prob.} & \textbf{Min. prob.} & \textbf{Min. finger}      & \textbf{Max. prob.} \\ \hline
					30                  & 0.82                & 0.72                & Little fingers and thumbs & 0.89                \\
					70                  & 0.79                & 0.70                & Little fingers            & 0.87                \\ \hline
				\end{tabular}
			\end{center}
			\bigskip
		\end{minipage}
	\end{table}
	
	\subsubsection{Touch Detection. }
	We recorded several videos pressing a surface to validate the performance of touch detection. Results are showed in Table~\ref{tab:touch processing module result}. Two index fingers and two middle fingers have high accuracy, meanwhile two ring fingers and two little fingers have high wrong probability, since they move a lot and they are not observed from a direct view.
	
	\begin{table}[t!]
		\caption{Touch detection result.}
		\label{tab:touch processing module result}
        \centering
				\begin{tabular}{|c|c|c|c|}
					\hline
					\textbf{Finger}     & \textbf{No. pressing time} & \textbf{True Positive} & \textbf{False Negative} \\
					\hline
					Right thumb         & 3                            & 3                      & 1                       \\
					Right index finger  & 9                            & 8                      & 1                       \\
					Right middle finger & 7                            & 5                      & 2                       \\
					Right ring finger   & 6                            & 3                      & 1                       \\
					Right little finger & 7                            & 3                      & 2                       \\
					Left thumb          & 3                            & 1                      & 0                       \\
					Left index finger   & 8                            & 6                      & 0                       \\
					Left middle finger  & 8                            & 5                      & 1                       \\
					Left ring finger    & 7                            & 4                      & 0                       \\
					Left little finger  & 6                            & 4                      & 3     \\            
					\hline  
				\end{tabular}
	\end{table}
	\section{User study}
	\label{section: User study}
	
	
	We invited participants to experience our system and share their feedback. The participants include 20 people (70\% of whom are male) with age between 10 and 51 (most of them are from 18 to 30.)
	
	We asked each participant to complete two tasks. The first task is to type a piece of text (using one key at a time). The second task involves formatting the text (using a combination of multiple keys such as bold, italic, copy, paste, etc.).
	
\begin{figure}[t!]
		\centering
		\includegraphics[width=\linewidth]{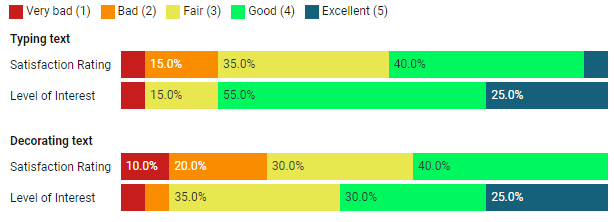}
		\caption{User study result.}
		\label{figure:user study results}
	\end{figure}

	After that, we surveyed users about several aspects of the system, including interest level and satisfaction (i.e., accuracy and smoothness). The scale ranges from 1 (very bad) to 5 (very good). The results presented in Figure~\ref{figure:user study results} shows that participants highly evaluated our proposed system in both tasks.
	
	Additionally, we gather user feedback on the system's limitations and what can be improved. The majority of users request improvements in smoothness and accuracy, as they sometimes experience lag in the system.
	
	\section{Conclusion}
	\label{section: conclusion}
	
	In this paper, we introduced a new approach to construct a low-cost virtual keyboard system by combining traditional computer vision algorithms and advancements in deep learning. We fine-tuned YOLOv8 models to detect and segment keyboard region and keyboard keys with high confident level, with the accuracy of 92\% for keyboard region, and 70\% for keys. We utilized the traditional algorithms to analyze the color of fingernail for touch detection with the accuracy about 36\%. \add{This result was not high since there are many factors affect the system (i.e. lighting conditions, camera angles, pressing force and fingernail of users).} We carried out a pilot study to obtain initial qualitative insights into the usability of our system. The findings demonstrated the advantages of the proposed system, particularly its efficiency in typing without the need for specialized equipment. Participants also offered useful feedback, pointing out areas for enhancement, including the need to improve the system’s smoothness and accuracy, as well as the suggestion that the application should have built-in guidance rather than relying on manual instructions.
	
	We have plan to enhance touch detection algorithms to achieve higher performance. We will combine OCR technique to classify keyboard layout better. Therefore we can fine-tune models for each keyboard layout. Moreover, we will integrate LLM models to auto-correct text while typing.
	
	\textbf{Acknowledgement. }  This research is supported by research funding from Faculty of Information Technology, University of Science, Vietnam National University - Ho Chi Minh City.
	%
	%
	%
	\bibliographystyle{splncs04}
	\bibliography{references}
\end{document}